\documentclass[aps,prd,amssymb,cite,
amsfonts,epsf,preprintnumbers,nofootinbib,superscriptaddress]{revtex4}

\usepackage[dvips]{graphicx}
\usepackage{bm,latexsym,amsmath,amssymb,amsfonts}
\usepackage[usenames,dvipsnames]{color}
\usepackage[colorlinks=true,linkcolor=blue]{hyperref}
\usepackage{color}
\usepackage{ulem}
\usepackage{epsfig}
\usepackage{braket}
\usepackage[mathscr]{eucal}
\usepackage{cancel}
\usepackage{mathrsfs}
\usepackage{pgf,tikz}
\usepackage{slashed}
\usetikzlibrary{arrows,automata}

\definecolor{mypink1}{rgb}{0.858, 0.188, 0.478}
\definecolor{mypink2}{RGB}{219, 48, 122}
\definecolor{mypink3}{cmyk}{0, 0.7808, 0.4429, 0.1412}
\definecolor{mygray}{gray}{0.6}
\definecolor{pptbg}{rgb}{0.961,0.945,0.863}

\newcommand{\be}[1]{\begin{equation} \label{#1}}
\newcommand{\ee}{\end{equation}}
\newcommand{\bea}{\begin{eqnarray}}
\newcommand{\eea}{\end{eqnarray}}
\newcommand{\ba}{\begin{array}}
\newcommand{\ea}{\end{array}}
\newcommand{\nn}{\nonumber}

\newcommand{\bel}{\begin{align}}
\newcommand{\eel}{\end{align}}

\newcommand{\ve}[1]{\vec{\bm{#1}}}
\newcommand{\Lie}{\pounds}

\newcommand{\hreff}[1]{\href{#1}{\color{blue}{#1}} }

\begin{document}
\title{Steady heat conduction in general relativity }

\author{Hyeong-Chan Kim}
\affiliation{School of Liberal Arts and Sciences, Korea National University of Transportation, Chungju 380-702, Korea}

\email{hckim@ut.ac.kr}



\begin{abstract}
We investigate the steady state of heat conduction in general relativity using a variational approach for two-fluid dynamics. 
We adopt coordinates based on the Landau-Lifschitz observer because it allows us to describe thermodynamics with heat, formulated in the Eckart decomposition, on a static geometry.
Through our analysis, we demonstrate that the stability condition of a thermal equilibrium state arises from the fundamental principle that heat cannot propagate faster than the speed of light. 
We then formulate the equations governing steady-state heat conduction and introduce a binormal equilibrium condition that the Tolman temperature gradient holds for the directions orthogonal to the heat flow.
As an example, we consider radial heat conductions in a spherically symmetric spacetime.
We find that the total diffusion over a spherical surface satisfies a red-shifted form, $J(r) \sqrt{-g_{tt}} =$ constant. 
We also discuss the behavior of local temperature around an event horizon and specify the condition that the local temperature is finite there.
\end{abstract}

\maketitle
 
\section{Introduction}

Astronomical observations indicate that most gravitating systems are non-static and radiative processes are vital mechanisms for energy dissipation. 
In the diffusion approximation, heat flux describes this process. 
Historically, Oppenheimer and Synder~\cite{Opp39} first addressed the problem of radiative gravitational collapse and presented a model based on a spherically symmetric dust cloud undergoing gravitational collapse. 
Later Vaidya~\cite{Vaidya51,Vaidtya52} derived the metric, which describes the exterior gravitational field of a radiating sphere. 
Then it became possible to model the interior of radiating stars by matching such solutions to the exterior Vaidya spacetime~\cite{Glass81,Santos85,Kramer92}.
Radiating models are also necessary for cosmology to describe phenomena like structure formation, the evolution of voids, and the study of singularities~\cite{Krasinski}.
Exact solutions for shear-free perfect fluids with heat flux were frequently studied to simplify the calculations and to allow realistic analytic solutions~\cite{Bonner89,Banerjee89,Govinder12,Ivanov12}. 
In these subjects, heat conduction plays a crucial role supported by thermodynamics. 

The studies of heat conduction in general relativity were performed for various purposes in various methods~\cite{IS1,IS2,IS3,Hiscock,Hiscock1987,Samuelsson:2009up,Andersson2011,Cesar2011,Andersson:2013jga,LK2022}. 
In the first era of research, researchers focused on the reconciliation between thermodynamics and general relativity.
Therefore, the primary topics were non-singularity, stability, and causality of thermodynamic systems. 
Most researches concentrated on analyzing thermal equilibrium and perturbations around it.
On the other hand, the action formulation developed by Taub~\cite{Taub54} and Carter~\cite{Carter72,Carter73,Carter89} allows one to deal with a highly non-equilibrium situation (usually called the ``off-the-shelf'' approach), at least in principle. 
The formulation was developed further to include dissipations and particle creations~\cite{Andersson:2013jga,AnderssonNew}. 

One of the most significant states in thermodynamics is the thermal equilibrium state. 
An important consequence of gravity in a thermal equilibrium state is the appearance of local temperature $\Theta(x^a)$, where $x^a$ with $a=0,1,2,3$ denotes a point in spacetime, usually called the Tolman temperature.
When the geometry is static, and matter keeps his position so that the comoving velocity $u^a$ becomes a unit vector along $ (\partial_t)^a$, the local Tolman temperature~\cite{Tolman} satisfies 
\be{Tolman}
\Theta(x^a) = \frac{\Theta_\infty}{\sqrt{-g_{tt}(x^a)}},
\ee
where $\Theta_\infty$ and $g_{tt}$ denote the temperature measured at a locally-flat asymptotic region and the time-time part of the metric for the corresponding geometry, respectively.
This formula holds even for stationary spacetimes~\cite{Buchdahl:49,Santiago:2018lcy} such as the Kerr-Newmann black holes.
Even when the geometry is not static, or matter does not follow a Killing trajectory, the Tolman temperature gradient for a system in thermal equilibrium still satisfies~\cite{Kim:2021kou}
\be{Tolman local}
\mathcal{T}_a =0, 
\ee
where the Tolman vector denotes
\be{Tolman vector} 
\mathcal{T}_a \equiv \frac{d(\Theta u_{a})}{d\tau} + \nabla_a \Theta   , \qquad \frac{d u_a}{d\tau} \equiv u^c\nabla_c u_a.
\ee
Here $u^a$ is the four-velocity of a local fluid element with $u_au^a =-1$, $\nabla_a$ is the covariant derivative with respective to the metric, and $\tau$ denotes the proper-time of an observer who moves with the velocity $u^a$.
Applying this relation to a system in thermal equilibrium in a static geometry, one may derive Eq.~\eqref{Tolman}.
Note that this equation presents three independent relations for directions normal to $u_a$ because the projection $u^a \mathcal{T}_a =0$ is an identity.
The stability of the thermal equilibrium state is also analyzed~\cite{Hiscock1987,Olson1990,LK2022}. 
For the thermal equilibrium state to be stable, the energy density $\rho(n, \sigma, q)$ must satisfy
\be{stability:TTES}
\left(\frac{\partial \rho}{\partial q}\right)_{n,\sigma} > \frac{q}{\rho +\Psi},
\ee
where $q$, $\sigma \equiv s/n$, and $\Psi$ denote the heat, the specific entropy(the entropy per unit particle), and the pressure, respectively.

In a recent work~\cite{LK2022}, we analyzed the heat-flow equation along the directions orthogonal to both the number and the entropy fluxes.
Then, we introduced a kind of zeroth law of thermodynamics that we call {\it binormal equilibrium condition} here. 
In this work, we stress the condition because it is not well known yet compared to its importance.
\vspace{.2cm}
\begin{figure}[htb]
\begin{center}
\begin{tabular}{c}
\includegraphics[width=.4\linewidth,origin=tl]{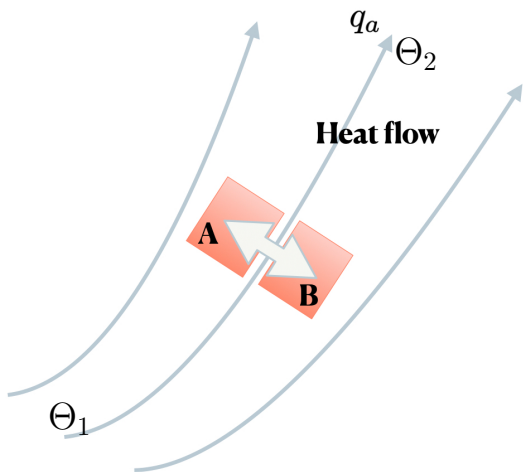} 
\end{tabular}
\put (-190,-69){\Large $\Theta_1$}
\put (-59,69){\Large $\Theta_2$}
\end{center}
\caption{The binormal equilibrium condition. 
Heat flows from a higher temperature region to a lower temperature region. ($\Theta_1> \Theta_2$)
The heat-flux vector $q_a$ lives in a 3-dimensional spacelike section perpendicular to $u^a$. Here, A and B denote two independent systems which are located along the direction perpendicular to the heat flux. }
\label{fig:Tol}
\end{figure}
In traditional thermodynamics without gravity, heat is directly linked to temperature difference.
Considering two neighboring systems, A and B, heat flows only when there is a temperature difference. 
Conversely, when heat flows between A and B, the temperature of A must be different from that of B. 
If the heat does not flow between the two as in Fig.~\ref{fig:Tol}, they are in equilibrium.
Here, we do not claim a thermal equilibrium because heat can flow along the direction perpendicular to the line connecting the two systems, since thermal equilibrium implies the absence of heat.
However, we can still say that the local temperatures of A and B are the same without loss of generality.
When gravity acts, thermal equilibrium is characterized by the Tolman temperature gradient~\eqref{Tolman local}. 
Therefore, the Tolman temperature gradient must be satisfied between subsystems A and B in the figure even if there is gravity.
In this sense, it is natural to require the Tolman temperature gradient to hold along the directions binormal to the particle trajectory and the heat flow: 
\be{Tolman local}
\perp_a^c \mathcal{T}_c =0, 
\ee
where $\perp_a^c$ denotes the projection operator both to the number and the entropy fluxes,
\be{perp}
 \perp_a^c \equiv  \delta _a ^c +u_a u^c -\frac{q_a q^c}{q^2} .
\ee 
Here, $q^a$ denotes the heat flux. 
Naturally, the heat $q$ becomes 
\be{q:q a}
q  \equiv \sqrt{q^a q_a}.
\ee
In Ref.~\cite{LK2022}, we also reformulated the relativistic analogy of the Cattaneo equation to reflect the {\it binormal equilibrium condition}~\eqref{Tolman local} by using the variational formulation of thermodynamics. 
The present article is based on the results, which we summarize in Sec.~\ref{Sec:II}. 
Recently, we showed that the action formulation allows the non-vanishing particle creation rate in a general setting when one considers the contribution to the variational formula appropriately~\cite{LK2022-2}. 
This condition fills the last piece of the thermodynamic equations in general relativity.

\vspace{.2cm}
Once we know the thermal equilibrium, it is natural to ask about the steady heat flow in general relativity.
In thermodynamics, heat conduction happens in a system located between two heat reservoirs of different temperatures. 
Because of the temperature difference, the system's temperature has gradients, which is a critical difference from the thermal equilibrium state.
As discussed in Ref.~\cite{Oono98,Sasa06}, where the authors tried to construct the steady state thermodynamics from statistical mechanics, non-equilibrium steady states are out of equilibrium but have no macroscopically observable time dependence.
Therefore, for a steady state of heat flow, all thermodynamic quantities including the intensity and the flow direction of heat $q$ may not change with time with respect to a specific coordinate system.
Naturally, the shape of the system is also invariant with respect to the time evolution to preserve the time independence of the thermodynamic quantities.
Therefore, the {\it steady-state condition} in general relativity becomes:
There exists a time-like Killing vector $\xi^a$ satisfying 
\be{stationary}
 \Lie_\xi q = 0, \qquad \Lie_\xi \rho=0, \qquad \Lie_\xi\Theta =0, \qquad \Lie_\xi \chi =0 ,
\ee 
where $\Lie_{\xi}$ denotes the Lie derivative with respective to the vector $\xi^a$ and 
$\chi$ denotes the chemical potential and the thermodynamic parameters are measured by a comoving observer with the vector $\xi^a$.
Detailed discussion on the steady state is given in Sec.~\ref{sec:SHF}. 
In general relativity, these requirements are nontrivial contrary to the traditional theory of thermodynamics. 
To argue this fact, we observe the stress tensor consisting of a thermal system. 
In general, the heat contributes to the off-diagonal part of the stress tensor with respect to a comoving reference frame. 
Therefore, when the comoving observer measures the time with his watch, he may find that the metric is not static but time-dependent. 
Because of the fact, we cannot use comoving coordinates to describe a steady state. 
For the choice of a coordinate system, we study in Sec.~\ref{sec:SHF}.
We also study the mild heat flow case for a steady state in Sec.~\ref{sec:IV} and summarize the results in Sec.~\ref{sec:V}.

\section{Heat conduction in a two-constituent model for relativistic thermodynamics} \label{Sec:II}

In this section, we survey the heat conduction equation, which is usually called the relativistic analogy of the Cattaneo equation, based on the action formalism for thermodynamics. 
The formalism was studied originally by Carter~\cite{Carter89}.  Later, it was improved by Priou~\cite{Priou1991} and Lopez-Monsalvo \& Andersson~\cite{Andersson11}.

The variational formulation of relativistic thermodynamics stems from the assumption that the matter flux $n^a$ and the entropy flux $s^a$ are two independent fluids interacting with each other.
The particle number in the system is assumed to be large enough that the fluid approximation is applied and there is a well-defined matter current $n^a$.  
As discussed in Ref.~\cite{Haskell2012}, it is the same as assuming that each constituent has a short enough internal length scale to perform averaging, while any mechanism that couples the flows acts on a larger length scale or on a longer time scale. 
A typical system of this kind is laboratory superfluids~\cite{Carter94,Andersson11}.
In this model, the entropy flux $s^a$ is, in general, not aligned with the particle flux $n^a$. 
The misalignment is associated with the heat flux, $q^a$, leads to entropy creation. 
The formulation is described in Eckart decomposition\footnote{ In this work, we use the term ``decomposition" instead of ``frame" because the latter is frequently used to call a frame of reference or a coordinate system.} where the observer's four-velocity $u^a$ is parallel to the number flux.
Explicitly, given the number density $n$, the entropy density $s$, and the heat flux $q^a$, the particle number and the entropy fluxes are 
\be{na}
n^a \equiv n u^a, \qquad s^a \equiv s u^a + \varsigma^a; \qquad \varsigma^a \equiv \frac{q^a}{\Theta},
\ee
where $q^a u_a =0$.
In this work, the heat flux $q^a$ denotes the deviation of the entropy flux relative to the number flux. 
This procedure defines the heat uniquely irrespective of the choice of coordinate system at least for this two-fluid model.

We now have $8$-independent unknown parameters $n$, $s$, $u^a$, and $q^a$.
The variation of a master function (Lagrangian) $\Lambda(n,s,\varsigma \equiv q/\Theta)$ gives 
$
\frac1{\sqrt{-g}}\delta (\sqrt{-g} \Lambda) = 
\chi_a \delta n^a + \Theta_a \delta s^a +\frac12 T^{ab} \delta g_{ab}
$ up to total derivatives, where $T_{ab}$, $g_{ab}$, and $g$ denote the stress tensor, the metric tensor, and its determinant, respectively.
Here, the conjugate momenta to $n^a$ and $s^a$ take the forms, 
\be{bar chi}
\chi_a = \mu \Theta_a + \nu q_a = \mu \Theta u_a 
	+\alpha q_a, \quad \Theta_a  
=  \Theta u_a + \vartheta_a ; 
 \qquad 
 \qquad \alpha \equiv  \mu\beta +\nu,
  \qquad \vartheta_a \equiv  \beta q_a, 
\ee
where $\Theta$ and $\chi \equiv \mu \Theta$ are the temperature and the chemical potential measured by a comoving observer with the number flux, respectively. 
Because the partial differentiations for $\Lambda$ commute, there exists a symmetry written as
$\Theta_{[a} s_{b]} + \chi_{[a} n_{b]} = 0$, which presents a relation between the variables, 
\be{ABC}
n \alpha + s\beta = 1 .
\ee
As shown in Ref.~\cite{LK2022}, the stability condition~\eqref{stability:TTES} for a thermal equilibrium state can be written in a simple form,
\be{stb}
\nu = \frac1n \left[1- \frac{\rho +\Psi}{\Theta} \beta \right] < 0,
\ee
which constrains the value of $\beta$ to be larger than $\Theta/(\rho+\Psi)$.
Here, we combine Eq.~\eqref{ABC} with the last but one of Eq.~\eqref{bar chi} and use $n\chi+s\Theta = \rho+\Psi$. 
Note that $-\Theta_a\Theta^a = \Theta^2
\left[1- \left(\frac{\beta q}{\Theta}\right)^2 \right]$ and  $-\chi_a\chi^a = \chi^2
\left[1- \left(\frac{\alpha q}{\chi}\right)^2 \right]$.
Because both $\Theta_a$ and $\chi_a$ are time-like vectors, the value of $|q|$ are bounded above by 
\be{q:bound}
q^2 \leq \mbox{min}\left( \frac{\Theta^2}{\beta^2}, ~ \frac{\chi^2}{\alpha^2} \right),
\ee
where $\mbox{min}(A,B)$ denotes a minimum value of the two.
The stress tensor, explicitly using Eqs.~\eqref{na}, \eqref{bar chi}, and \eqref{ABC},  becomes
\be{stress tensor}
T_{ab} = \Theta_a s_b+ \chi_a n_b +\Psi g_{ab} 
= \left(\frac{\beta q^2}{\Theta}-\Lambda\right) u_a u_b + (u_a q_b+u_b q_a) +\Psi \gamma_{ab} + \frac{\beta}{\Theta} q_a q_b ,
\ee
where $\gamma^{ab} \equiv g^{ab} + u^a u^b$ denotes the projection normal to the vector $u^a$ and $\Psi$ denotes the pressure,
\be{pressure}
\Psi = \Lambda - \Theta_a s^a -\chi_a n^a = \Lambda+\Theta s +\chi n -\frac{\beta q^2}{\Theta}.
\ee
The energy density with respect to the comoving observer is
\be{density}
\rho(n,s,\vartheta) \equiv u^a u^b T_{ab}  
	= \vartheta \varsigma -\Lambda(n,s,\varsigma)
	=\frac{\beta q^2}{\Theta}-\Lambda  .
\ee
Note that this equation presents a Legendre transform from $\Lambda(n,s,\varsigma)$ to $\rho(n,s,\vartheta)$.
Therefore, one gets the variational law of $\rho$ from that of $\Lambda$, which presents the first law of thermodynamics.
Note also that, to this comoving observer, the off-diagonal element of the stress tensor defines the heat:
\be{heat:comoving}
q^c =- u^a\gamma^{bc} T_{ab} ,
\ee

Therefore, the heat flux $q^a$ is defined by the energy flux measured by a comoving observer with the matter. 

\vspace{.2cm}
The theory of heat conduction in Ref.~\cite{LK2022} consists of the particle/entropy creation relations, two heat-flow equations, and the binormal equilibrium condition~\eqref{Tolman local}. 
The two heat-flow equations consist of two differential equations:
One is the relativistic analog of the Cattaneo equation and the other (new equation) originates from the $q^a$ part of the energy-momentum conservation equation. 

Let us write the heat-flow equations starting from their binormal parts to both $u^a$ and $q^a$.
Because we live in four dimensions, there are four independent binormal degrees of freedom for the evolution of $u^a$ and $q^a$.
To describe the evolution, therefore, we need four independent equations.
Two equations come from the binormal equilibrium condition~\eqref{Tolman local}.
The other two come from the binormal part of the relativistic analogy of the Cattaneo equation Eq.~(48) in \cite{LK2022}\footnote{In this work, we write the equation in a slightly modified form from that in Ref.~\cite{LK2022}. We separate the binormal direction equation from those along the heat. We also use the binormal equilibrium condition~\eqref{Tolman local} to simplify the equation.}:
\be{eq:bi}
      \perp_a^c \left[ \frac{d q_c}{d\tau} + (\nabla_c u_b) q^b +\frac{ 
      \left[ \ve{q}\cdot \bm{d\Theta}\right]_c 
}{ \Theta} \right] 
= - n\Theta  \perp_a^b\nabla_b \mu   ,
\ee
where, later in this work, we use
\be{q d theta}
[\ve{q}\cdot \bm{d\Theta}]_b = u_b (q^c\nabla_c \Theta) + \Theta q^c(\nabla_c u_b - \nabla_b u_c) + q^c (\nabla_c \vartheta_b -\nabla_b \vartheta_c).
\ee
The combination of Eq.~\eqref{Tolman local} and \eqref{eq:bi} determines the binormal-directional behaviors of a thermal system. 

There remains another four-independent equations which describe the behaviors along $u^a$ or $q^a$.
Two of them are nothing but the particle and the entropy creation equations, 
\be{creation}
\Gamma_n = \nabla_a n^a =0, \qquad \Gamma_s = \nabla_a s^a = \frac{q^2}{\kappa \Theta^2},
\ee
where $\kappa \geq 0$ denotes the thermal conductivity. 
Dissipations such as shear or viscosity present positive contributions to the right-hand side of Eq.~\eqref{creation}.  
Here, we choose the number creation rate to vanish for the steady heat flow state.
The other two evolution equations come from the $q^a$ parts of the relativistic analogy of the Cattaneo equation and the energy-momentum conservation equation.
The first is 
\be{cat2}
\left(1+\gamma_n + \kappa \frac{d \beta}{d\tau}\right) \frac{q}{\kappa} 
    + \beta \left[ \hat q^a\frac{d q_a}{d\tau} + (\nabla_a u_b) \hat q^aq^b\right] 
= -\hat q^b\mathcal{T}_b ,
\ee
where $\hat q^a \equiv q^a/q$ is a unit vector along the heat flux and $\gamma_n \equiv \kappa \Theta \chi \Gamma_n/q^2$ is a dimensionless combination of the number creation rate.
Here, we get this equation by multiplying $q^a$ on both sides of the relativistic analogy of the Cattaneo equation, Eq.~(48) in Ref.~\cite{LK2022}.
The remaining equation is the $q^a$ part of the energy-momentum conservation equation (Eq.~(50) in \cite{LK2022}):
\be{eq2}
 \left[ \sigma-\frac{\beta q^2}{n\Theta^2}
	+ \left(\sigma -\frac{\alpha q^2}{n \Theta \chi} \right)\gamma_n
	-\kappa  \frac{d\alpha}{d\tau}
    \right]\frac{q}{\kappa} - \alpha
\left[ \hat q^a \frac{dq_a}{d\tau} + (\nabla_{a} u_{b}) \hat q^aq^b \right]
 = \hat q^b \mathcal{K}_b ,
\ee
where 
$$
\mathcal{K}_b \equiv \frac{d(\chi u_b)}{d\tau} + \nabla_b \chi.
$$
The right-hand side of Eq.~\eqref{eq2} is nothing but the heat-directional component of $\mathcal{K}_b$. 
Note that the left-hand sides of Eqs.~\eqref{cat2} and \eqref{eq2} vanish when $q \to 0$. 
Therefore, taking the $q\to 0$ limit, the equations present $\hat q^b \mathcal{T}_b=0=\hat q^b \mathcal{K}_b$.
Because the direction of heat is meaningless in the limit except that it is spatial, this relation automatically implies $\gamma_u^{ab}\mathcal{T}_b=0=\gamma_u^{ab} \mathcal{K}_b $.
These equations are nothing but  the Tolman temperature gradient and Klein's relation\footnote{As noted in Ref.~\cite{Kim:2021kou}, Klein's relation may not hold for models with more than three fluids.}~\cite{Klein49} for thermal equilibrium systems, respectively.

\section{ Steady heat flow } \label{sec:SHF}

When studying heat flow in thermodynamics, one considers a system located between two heat baths of different temperatures. 
The temperature difference (at the present case $\mathcal{T}_a \neq 0$) distinguishes the steady state from the thermal equilibrium state. 

Before delving into the general relativistic case, it is important to clarify the meaning of the term ``steady state" in non-equilibrium thermodynamics. 
A steady state is reached when the system has relaxed to a stationary regime, and transient effects can be ignored when describing heat conduction. 
A non-equilibrium steady state refers to a macroscopic physical system that exchanges energy continuously with its environment but shows no observable macroscopic changes~\cite{Oono98,Sasa06}. 
This definition applies universally across various fields of study, including thermodynamics, fluid dynamics, chemistry, and engineering. 
Thus, in a steady state, all thermodynamic quantities, including local temperature and the shape of the heat baths, remain constant over time.

In this section, we elucidate the concept of a steady state in general relativity. 
For systems in a steady state, all thermodynamic parameters remain constant in time.
Specifically, in the Eckart decomposition, we consider a collection of matter whose comoving vector is denoted by $u^a$.
Let heat flows along a radial direction. 
It appears natural to choose the time coordinate to be generated by the vector $u^a$ and one of the space coordinates (e.g., radial direction) to be parallel to the heat.
With respect to an observer in this coordinates, matter appears static, and the heat manifests as momentum flux in the stress tensor. 
Considering a general spherically symmetric geometry, the time-space component of the Einstein tensor comes from the time dependence of the metric functions rather than an off-diagonal part of the metric.
This fact implies that the Eckart coordinates is not appropriate to describe a steady state of heat flow because the steady-flow condition requires the geometry to be independent of time similarly to the shape of the heat baths in the previous example.

On the other hand, there are a set of observers comoving with a timelike unit vector $v^a$ who notice the local momentum density vanishes. 
In this work, we call them the Landau-Lifschitz (LL) observers~\cite{LL}.
To make the local momentum density vanishes, the LL observers moves along the direction of the heat from the viewpoint of the Eckart observers.
Conversely, with respect to LL observers, the matter flows in the opposite direction to the heat, canceling out the momentum of heat and resulting in a zero total momentum density. 
Incorporating the time coordinate for the LL observers, one can construct a coordinates frame. 
Then, the stress tensor does not have an off-diagonal component, allowing the geometry to be locally time-independent in this coordinates.
The coordinate choice will be dependent on the global situations because the LL frame is locally equivalent to the Eckart frame. 
For our purposes, we incorporate the LL coordinates since it allows us to use the time-independence for a steady state with the steady-state condition in Eq.~\eqref{stationary}.
Notice that, the existence of a steady state depends on  the existence of (locally) static coordinates.
While, thermal equilibrium state does not presume any geometrical property but is determined from the thermodynamic properties only.

Next, we examine the implications of choosing the LL coordinates. 
To do so, we consider a system with two thermal baths attached at either end, each with a different temperature. 
Since we are dealing with a steady state, both baths must remain stationary in the LL coordinates. 
However, this does not mean that they are stationary with respect to the comoving observer with the matter. 
Due to the simultaneous flows of both heat and matter, the thermal baths must allow not only heat flow but also diffusion of matter.
Here, the heat flow must be related to the diffusion in such a way that ensures the total momentum density vanishes.

In an ordinary star, matter does not move or moves slowly, with only heat propagating out. 
As a result, the momentum flow does not vanish to a comoving (Eckart) observer with matter, rendering at least one of the steady state assumptions in Eq.~\eqref{stationary} invalid. 
Consequently, the thermodynamic quantities vary gradually in such a star. 
However, this does not mean that the concept of a steady state is useless in understanding such stars. 
The slow changes can be described by a successive sequence of steady states, similar to a quasi-static change by that of static ones.
We can envisage an idealized system in which heat flows steadily. 
For instance, consider a binary star consisting of a neutron star and a massive giant. 
Typically, matter in the giant is slowly absorbed into the neutron star through the accretion disk around it. 
If this absorption process is slow and lasts long enough, we can consider the absorbing matter in the accretion disk as a thermal system in a steady state, with matter flowing from the giant to the neutron star and heat flowing out. 
Hence, we restrict ourselves to studying steady states in this work.

\vspace{.1cm}
When a thermodynamic system is in a steady state, there exists a timelike Killing vector $\xi^a$ along which the geometry does not change:
$$
\Lie_{\xi} g_{ab} =0,
$$
at least in the region of the our interest.
By normalizing the Killing vector, we define a unit vector $v^a \equiv \xi^a/|\xi|$  which generates the time.
With the frame of reference using this time, the momentum density vanishes: 
\be{LL}
v^a(g^{bc} + v^b v^c) T_{ac}=0.
\ee 
In this sense, this vector $v^a$ defines the Landau-Lifschitz (LL) coordinates.
Note, however, that the absence of energy flux does not imply the absence of heat, $q$.
The reason is that the energy flux consists of the heat flux and the number flux\footnote{In some literature~\cite{Tsumura:2012ss}, heat is identified with the energy flux itself. }. 
The number flux flows along the opposite direction to the heat flux so that their energy flows cancel each other.
In other words, while we work in a LL coordinates, heat is defined by Eq.~\eqref{heat:comoving} in a comoving frame.
When a system is in a steady state, the thermodynamic system stays in a `static' state in a sense that the geometry, the energy density, the temperature, the chemical potential, and the heat do not change with time satisfying Eq.~\eqref{stationary}. 
Of course, these equations do not imply that the fluids are static.
The entropy flux flows with the opposite directions to the matter flux so that the total energy flux vanishes.

\vspace{.3cm}
Because the thermodynamic equations are written in the Eckart frame and the steady state is defined in the LL coordinates, we need to find a relation between the two. 
The relation comes from the transformation between the comoving velocities, $u^a$, and the time-generating vector, $v^a$.  
Locally, the four-vector $v^a$ must be a linear combination of $u^a$ and $q^a$. 
Let us introduce a unit vector $\hat q^a$ along the heat that satisfies $g_{ab}\hat q^a \hat q^b =1$. 
Because both $u^a$ and $v^a$ are unit future-directed timelike vectors, we have
\be{u v}
v^a =  \cosh \epsilon \, u^a + \sinh \epsilon \, \hat q^a , \qquad 
\hat j^a = \sinh \epsilon \,u^a +  \cosh \epsilon \,\hat q^a.
\ee
Here $\hat j^a$ denotes the unit-spacelike vector along the heat flux normal to $v^a$ with $g_{ab} \hat j^a \hat j^b=1$ and $\epsilon$ is a function of thermodynamic quantities to be determined by the condition for the LL frame~\eqref{LL}, respectively.
Therefore, $\tanh \epsilon$ denotes the velocity of the LL observer relative to the Eckart observer.
Inverting the equation, we have
\be{v u}
u^a = \cosh \epsilon \, v^a - \sinh \epsilon \, \hat j^a, \qquad 
\hat q^a = - \sinh \epsilon\, v^a + \cosh \epsilon \, \hat j^a .
\ee
With respect to the LL observer, the number and the entropy fluxes become, from Eq.~\eqref{na},
\be{ns}
n^a = n \cosh \epsilon \, v^a - n \sinh \epsilon \, \hat j^a, \qquad
s^a = \left(s \cosh \epsilon-\frac{q\sinh\epsilon}{\Theta}\right) v^a
	+ \frac{q}{\Theta} \left(\cosh \epsilon -\frac{\Theta s \sinh \epsilon}{q} \right) \hat j^a.
\ee
Therefore, the number flux has diffusion, 
\be{diffusion}
 -n \sinh \epsilon,
\ee
along the  direction opposite to the heat.
In this sense, the number and the entropy fluxes are not static  with respect to the LL observer.
However, for the system be steady in time, we require the number and the entropy densities with respect to the observer $v^a$, 
\be{bar n}
\bar n \equiv -v^a n_a = n\cosh \epsilon , \qquad
\bar s \equiv - v^a s_a =  s\cosh \epsilon\left(1 - \frac{q}{\Theta s} \tanh \epsilon \right),
\ee
do not change with time.
Here the barred quantities denote the corresponding quantities with respect to a LL observer. 
Because the heat $q$ also do not change with time for a steady state, we may set all the physical parameters $n$, $s$, and $q$ to be independent of time. 
Note that the number density with respect to the LL observer is higher than that of the Eckart observer for all $\epsilon$. 

The temperature and the chemical potential covectors become 
\bea
\Theta_a &=& \Theta u_a + \beta q_a 
= \Theta \cosh \epsilon \left[\left(1- \frac{\beta q}{\Theta} \tanh \epsilon \right) v^a 
	+ \left(\frac{\beta q}{\Theta} - \tanh \epsilon\right) \hat j^a \right], \nn \\
\chi_a &=& \chi u_a 
	+\alpha q_a
= \chi \cosh \epsilon\left[
	\left(1 - \frac{\alpha q}{\chi} \tanh \epsilon \right) v^a 
	+ \left(\frac{\alpha q}{\chi}  -  \tanh \epsilon\right) \hat j^a
	\right] .
\label{Theta chi}
\eea
Therefore, the temperature and the chemical potential with respect to the LL observer are
\be{b Theta}
\bar \Theta \equiv -v^a \Theta_a = \Theta \cosh\epsilon \left(1-\frac{\beta q}{\Theta} \tanh \epsilon \right), \qquad
\bar \chi \equiv \chi \cosh \epsilon \left(1 - \frac{\alpha q}{\chi} \tanh \epsilon \right) ,
\ee
Note that the non-negativity of $\bar \Theta$ and $\bar \chi$ are guaranteed by the inequalities in Eq.~\eqref{q:bound}.

Now, we calculate the energy flux~\eqref{LL} from the stress tensor~\eqref{stress tensor} by using the above equations:
\bea
- v^a \hat j^b T_{ab} 
&=&-\varepsilon \sinh(2\epsilon) 
	+ q\cosh(2\epsilon); 
\qquad \varepsilon \equiv \frac{\beta q^2}{\Theta} +\frac{\Psi- \Lambda}2 
	= \frac12\left[ \rho +\Psi +\frac{\beta q^2}{\Theta}\right] ,
\eea
where we use Eq.~\eqref{density}.
Now, the LL frame is defined by the condition that $- v^a \hat j^b T_{ab} =0$ from Eq.~\eqref{LL}. 
This condition determines $\epsilon$ to satisfy
\be{epsilon q}
\tanh 2\epsilon = \frac{q}{\varepsilon} ,
\ee
where $\tanh\epsilon$ denotes the relative velocity between $v^a$ and $u^a$.
When heat is small enough, $q/(n\Theta) \ll 1$, the relation~\eqref{epsilon q} gives 
\be{epsilon:q}
\epsilon \approx\frac{q}{\rho+\Psi}
	      \approx \frac{q}{n\Theta (\mu+\sigma) } \ll 1
\ee
for $\rho \neq -\Psi$.
Note also that the right-hand side of the equality in Eq.~\eqref{epsilon q} satisfies, for all real values of $q$, 
\be{stb1}
\frac{|q|}{\varepsilon} =  2\left[(\mu+\sigma)\frac{n\Theta}{|q|} +n\beta \frac{ |q|}{n \Theta}\right]^{-1} \leq \frac{1}{\beta n (\mu+\sigma)} ,
\ee
where the positivity of $\beta$ and $\mu+\sigma$ are assumed.
The left-hand side of the equality in Eq.~\eqref{epsilon q}, $\tanh 2\epsilon$, denoting the relative velocity of the energy fluxes of heat and the number is not larger than one for any real number $\epsilon$.
When $\beta n (\mu+\sigma) < 1$, the right-hand side can be larger than one for some $q$. 
In this case, the corresponding value of $\epsilon$ is not real for that $q$.
This result is discrepant from the fact that the two unit vectors $u^a$ and $v^a$ are timelike and any two timelike unit vectors can be converted to each other by a (local) Lorentz transform.
To avoid this case, one needs to constrain $\beta n (\mu+\sigma) > 1$.
Note that this constraint is equivalent to the stability condition in Eq.~\eqref{stb} for a thermal equilibrium state.
This result is an interesting justification for the stability condition of  thermal equilibrium systems.

Now, the energy density and the pressure along the $\hat j^a$ direction with respect to the LL observer become
\bea 
\bar \rho &=& v^a v^b T_{ab} 
=\varepsilon \cosh 2\epsilon 
 	-q\sinh(2 \epsilon) -\frac{\Psi+\Lambda}{2} , \nn \\
\bar P &=& \hat j^a \hat j^b T_{ab}
=\varepsilon \cosh 2\epsilon
	+ \frac{\Psi +\Lambda}2
	- q \sinh (2\epsilon) 
.  \label{rho P}
\eea
Interestingly, the sum of the energy density and pressure with respect to the LL observer satisfies
\be{rho+P}
\bar \rho + \bar P 
=2 \cosh(2\epsilon) \left( \varepsilon
	- \frac{q^2}{ \varepsilon} \right) = 2\sqrt{\varepsilon^2- q^2} \geq 0.
\ee
This value vanishes only when the stability condition satulates the equality in Eq.~\eqref{stb1} with $\beta n (\mu+\sigma) =1$.

\subsection{Radial steady heat flow in a spherically symmetric spacetime}
One may analyze part of the steady state condition~\eqref{stationary} without introducing  an explicit coordinate system.
However, the other parts require explicit coordinates and paths of the matter and heat.
In this work, we consider a simple configuration of a steady heat flow system, a radial heat flow along $\hat r$ direction in a spherically symmetric geometry.
Therefore, $\hat j^a \propto (\partial_r)^a $, which gives $\hat j^a = \hat r^a$. 
With the choice $v^a \propto (\partial_t)^a$, the LL coordinates is described by the static and spherically-symmetric metric,
\be{metric1}
ds^2 = -e^{2N} dt^2 + f^{-1} dr^2 + r^2 d\Omega^2_{(2)} ;
\qquad -g_{tt} = e^{2N} \equiv f(r) e^{-h(r)},   \qquad r_- \leq r \leq r_+,
\ee
where $d\Omega^2_{(2)}$ denotes the metric of a unit sphere and all the metric functions are independent of time.
Here, we assume that the thermal system is within a spherical shell from $r_-$ to $r_+$. 
We are not interested in the geometry and physics outside the region.
Usually, reservoirs of heat will be located both inside, $r< r_-$, and outside, $r> r_+$, the shell. 
Let us examine the Einstein equation for the metric~\eqref{metric1}. 
The time-time part of the Einstein equation, $G_{tt} = 8 \pi \bar \rho$, gives 
$$
f(r) = 1-\frac{2M}{r}, \qquad M(r) = 4\pi \int^r {r'}^2 \bar \rho(r') dr' .
$$
Because $G_{tr} \propto d M/d\tau$ and $T_{tr} =0$, the mass $M$ should be independent of time, which matches the steady state condition. 
The other component of Einstein equation  gives $G_{tt} + G_{rr} = - fh'/r = 8  \pi (\bar \rho+\bar P)$, where the prime denotes the derivative with respect to $r$.
Writing the results with respect to $N$, we have
\be{N'}
N' = \frac{(\sqrt{-g_{tt}})'}{\sqrt{-g_{tt}} } = \frac{4\pi r}{f} ( \rho _0 + \bar P) ; \qquad \rho_0= \frac{M}{4\pi r^3}.
\ee
With this metric, the unit vectors $v^a$ and $\hat j^a$ become
\be{unit vectors}
v^a = ( e^{-N}, 0, 0, 0), \qquad \hat j^a = ( 0, \sqrt{f}  , 0, 0 ).
\ee
For later use, let us calculate the expansion, $ \nabla_a u^a$, of the matter paths when the heat flows along the radial direction in this coordinate system: 
\bea
\nabla_au^a &=& \frac1{\sqrt{-g}} \partial_a \left(\sqrt{-g} u^a\right) 
= \frac1{\sqrt{-g}} \partial_t[ \sqrt{-g} e^{-N}\cosh \epsilon]
- \frac1{\sqrt{-g}} \partial_r [ \sqrt{-g} \sqrt{f} \sinh\epsilon ] \nn \\
&\simeq &
-\sqrt{f} \sinh\epsilon\left[ \log (e^N r^2 \sinh \epsilon)\right]'
 , \label{expansion}
\eea
where the equality ``$\simeq$" in the second line denotes that we use the steady state approximation.
We also calculate $\nabla_a(q^a/\Theta)$ based on the state state approximation, 
\bea
\nabla_a \frac{q^a}{\Theta} &=& \frac1{\sqrt{-g}} \partial_a \left( \frac{\sqrt{-g} (- q \sinh \epsilon \, v^a + q\cosh\epsilon\, \hat j^a)}{\Theta} \right) \nn \\
&=& - \frac1{\sqrt{-g}} \partial_t \left( \frac{\sqrt{-g} e^{-N} q \sinh \epsilon  }{\Theta} \right) 
+\frac1{\sqrt{-g}} \left( \frac{\sqrt{-g} \sqrt{f} q\cosh\epsilon}{\Theta} \right)'  \nn \\
& \simeq& \frac{\sqrt{f} q \cosh \epsilon}{\Theta}
 \left(\log \frac{e^N r^2 q\cosh \epsilon}{\Theta}\right)'  .  \label{d qT}
\eea

\vspace{.3cm}

Now, we are ready to analyze the equation of motions for  for spherically symmetric, steady, radial heat flow. 
Let us begin with the binormal parts. 
\begin{enumerate}

\item Let us consider the binormal equilibrium condition~\eqref{Tolman local} first.
Because the acceleration $ du_a/d\tau$ has only components along $q^a$ for the radial heat flow, we find  
\be{physicality}
\perp_a^b \mathcal{T}_b = \perp_a^b \nabla_b \Theta =0 \quad
\rightarrow \quad \partial_k \Theta =0, \qquad k = 2,3.
\ee
This equation implies that the local temperature should be the same on a spherical shell of the same radius, even if there is  radial heat. 

\item  
Next, we consider the binormal part of the relativistic analogy of the Cattaneo equation~\eqref{eq:bi}.
For a radial heat flow, $dq_c/d\tau$, $(\nabla_cu_b)q^b$, and $[\ve{q}\cdot \bm{d\Theta}]_c$ do not have components along the angular directions\footnote{Explicitly, one may use Eq.~(40) in Ref.~\cite{LK2022}. There, $\perp_a^c [\ve{q}\cdot \bm{d\Theta}]_c$ term is given by the linear combination of $Q_a^\perp$ and $\tilde Q_a^\perp$.
In thermal equilibrium, both $Q_a^\perp$ and $\tilde Q_a^\perp$ vanish  and the subsystems located along the binormal directions are in equilibrium.}.
This fact makes the left-hand side of Eq.~\eqref{eq:bi} vanish.
Therefore, we get 
\be{bi Cat}
\perp_a^b \nabla_b \mu  =0 \quad \rightarrow \quad \partial_k\mu =0, \qquad k= 2,3.
\ee
This equation indicates that the local chemical potential to temperature ratio should be the same on the spherical shell of the same radius. 
\end{enumerate}
Because $q$, $\Theta$, and $\chi$ are independent of angular coordinates, all the physical quantities will be independent of the angular coordinates for this radial heat flow, which is consistent with the spherical symmetry.
Note that these two equations~\eqref{physicality} and \eqref{bi Cat} are equally held even for thermal equilibrium states. 
The results are related to the fact that no heat flows along the binormal directions. 

\vspace{.1cm} 
Next, we consider the heat conduction equations along the directions $u^a$ and $q^a$.
In general, the equations are mixed together and form a coupled equations for $n$, $s$, and $q$.
\begin{enumerate}
\item
Let us begin with the particle number conservation equation, the first equation in Eq.~\eqref{creation}.
By using Eq.~\eqref{bar n},
\bea \label{dn}
0= \Gamma_n &=& u^a \nabla_a n+n  (\nabla_a u^a) 
  \nn \\
 &=& (v^a \nabla_a\bar n) -n(\sinh\epsilon) \, (v^a \nabla_a \epsilon)  - (\sinh \epsilon) (\hat j^a \nabla_a n)  + n (\nabla_a u^a) .
\eea
The local number density $\bar n$ must be invariant for the steady state with respect to the LL observer moving with $v^a$. 
This gives, 
$$
(v^a \nabla_a \bar n) =(n \sinh \epsilon) \left[(v^a\nabla_a \epsilon) +\frac{(\hat j^a \nabla_a n)}{n } -\frac{ (\nabla_a u^a) }{\sinh \epsilon}\right] =0.
$$
Because both $q$ and $\varepsilon$ are functions of thermodynamic quantities, we require that $\epsilon$ is independent of time from Eq.~\eqref{epsilon q}, $v^a\nabla_a \epsilon =0$, for a steady heat flow.
Therefore, we get 
\be{n'}
\frac{(\hat j^a \nabla_a n) }{n} = \frac{ (\nabla_a u^a) }{\sinh \epsilon} .
\ee
The expression until now holds irrespective of coordinate choice.

We now express the particle number conservation equation~\eqref{n'} for a radial heat flow for the spherically symmetric coordinates~\eqref{metric1} by using the expansion~\eqref{expansion} and Eq.~\eqref{unit vectors},
\be{n ep}
\frac{d \log (n \sinh \epsilon)}{dr} = -\left(\frac2r +N'\right).
\ee
Note that this equation can be integrated to present a position independent quantity $J_\infty$ for a steady state, 
\be{id1}
J_\infty \equiv  \sqrt{-g_{tt}} J(r) = \mbox{ position independent}; \qquad J(r) \equiv -4\pi r^2 (n^a \hat j_a) = 4\pi r^2 n \sinh \epsilon .
\ee
Here, $-J(r)$ denotes the sum of diffusion over a spherical surface of radius $r$.
For a steady heat conduction, this value satisfies
$
J_\infty(r_1) = J_\infty(r_2) =J_\infty
$
when $r_- \leq r_1, r_2\leq r_+$.
Because $\epsilon $ is a function of $q/\varepsilon$, this equation determines $q/\varepsilon$ as a function $\eta(r)$ over $n$:
\be{id2}
n \sinh\epsilon =\eta(r); \qquad \eta(r) \equiv \frac{J_\infty}{4\pi r^2 \sqrt{-g_{tt}}} = \frac{J_\infty}{4\pi r^2 e^N}.
\ee
From Eq.~\eqref{epsilon q}, we get  
\be{q:Q}
\frac{J_\infty}{4\pi r^2\sqrt{-g_{tt}} n}= \frac{\eta(r)}{n} = \frac{  q/\varepsilon}{\sqrt{2}(1-q^2/\varepsilon^2)^{1/4}
	\sqrt{1 +\sqrt{1-q^2/\varepsilon^2}}}.
\ee
Note that the right-hand side of Eq.~\eqref{q:Q} monotonically increases from negative infinity to infinity as $q/\varepsilon$ varies from minus one to one.
Therefore, for any $\eta/n$, a unique value of $q/\varepsilon$ will be specified.
Conversely, given the function $\eta(r)$, the heat becomes
\be{q:ep:eta}
\frac{q}{\varepsilon} \equiv \tanh 2\epsilon 
=\frac{ 2\sqrt{1 + (\eta/n)^2}}{ 1 + 2(\eta/n)^2} \frac{\eta}{n} .
\ee
When the system is in thermal equilibrium, Eq.~\eqref{dn} presents $dn/d\tau= -(\nabla_a u^a)n$.
Therefore, the number density becomes dynamic inevitably when $\nabla_a u^a \neq 0$.
This equation clearly shows the difference between a thermal equilibrium state from a steady state. 
The thermal equilibrium state allows time-dependent changes in thermodynamic parameters because it concerns only coordinate-independent properties. 
On the other hand, a steady state condition constrains the metric time dependence also.

\item
Next, we consider the second law of thermodynamics, $\Gamma_s = \nabla_a s^a = q^2/\kappa \Theta^2 \geq 0$.
Explicit calculation by using Eq.~\eqref{dn} gives 
\bea \label{ds}
 \frac{q^2}{\kappa \Theta^2}
=\Gamma_s 
	= \frac{d s}{d\tau} + s \nabla_a u^a 
		+ \nabla_a\left(\frac{q^a}{\Theta}\right)
	&=& \frac{d s}{d\tau}- \sigma\frac{d n}{d\tau}+ \nabla_a\left(\frac{q^a}{\Theta}\right)   \nn \\
\rightarrow s \frac{d}{d\tau} \log \sigma  
&=&\frac1{\kappa}
		\left(\frac{q}{\Theta} \right)^2
	- \nabla_a\left(\frac{ q^a}{\Theta}\right) 
 ,
\eea
where $\sigma \equiv s/n$ and $\mu \equiv \chi/\Theta$ are the specific entropy and the chemical potential to temperature ratio, respectively. 
Now, the left-hand side becomes
\be{ds1}
s \frac{d\log \sigma}{d\tau} = n u^a \nabla_a \sigma 
\simeq -(n  \sinh \epsilon) \,\hat j^a
	\nabla_a \sigma  ,
\ee
where we use $v^a \nabla_a  \sigma \simeq 0$ for a steady state. 
Putting Eqs.~\eqref{ds1} and \eqref{d qT} to Eq.~\eqref{ds}, the second law of thermodynamics gives
\be{sigma'}
- \sigma' = \frac{q^2}{(n \sinh \epsilon) \sqrt{f} \kappa \Theta^2} 
	- \frac{q \coth \epsilon}{n\Theta} 
	\left[ N' + \frac2r+\frac{n'}{n}+\left(\log\frac{q}{n\Theta}\right)' 
	+ \left(\log (\cosh\epsilon)\right)'\right].
\ee
Now, we use Eqs.~\eqref{n ep}, \eqref{app:qT}, and \eqref{app:chep'} to reduce the equation into the form,
\be{sigma':2}
 \frac{1+\eta^2/n^2}{1+2\eta^2/n^2} 
	 \left(\frac{2\varepsilon}{n\Theta}\right)' - \sigma' = \frac{q^2}{\eta \sqrt{f} \kappa \Theta^2} 
	- \frac{2\varepsilon}{n\Theta}
	 \frac{\frac{2\eta^2}{n^2}}{\left(1+\frac{2\eta^2}{n^2}\right)^2} 
		\left( \frac{n'}{n} 	
	 +N'+\frac2r\right) .
\ee
In this work, we prefer this form to Eq.~\eqref{sigma'} because, to the first order in $q$, the left-hand side becomes $\mu'$ from $2\varepsilon/n\Theta = \mu+\sigma+\beta q^2/n\Theta^2$ and $\eta \sim O(q)$. 
Note that the $O(q)$ term on the right-hand side of this equation is the first term, which originates from the entropy creation rate. 
Therefore, this result signifies that the chemical potential to the temperature ratio, $\mu$, grows in the presence of outgoing heat.

\item  We next consider the $q^a$ part of the relativistic analogy of the Cattaneo equation~\eqref{cat2}. 
For the calculation of the left-hand side, we need 
$$
\hat q^a \frac{dq_a}{d\tau} =  -\sqrt{f}q' \sinh \epsilon , \qquad 
(\nabla_a u_b)\hat q^b \hat q^a = - (N' \sinh \epsilon + \epsilon' \cosh \epsilon) \sqrt{f},
$$
which comes from the results in Eq.~\eqref{app:dq/dtau} and \eqref{app:d u q}.
The $q^a$ part of the relativistic analogy of the Cattaneo equation~\eqref{cat2} becomes, for a steady state,
\be{Tol2-1}
\hat q^b\mathcal{T}_b 
= -\frac{q}{\kappa} + \beta \sqrt{f}q \sinh\epsilon \left(\log \frac{\beta q}{nr^2}\right)',
\ee
where we use Eq.~\eqref{n ep}.
The left-hand side of this equation becomes, from Eq.~\eqref{Tolman q} for the metric~\eqref{metric1}, 
$$ 
\hat q^b\mathcal{T}_b=\sqrt{f} e^{-N} (\Theta e^N \cosh \epsilon)' .
$$
Putting this, we get
\be{Tol2-2}
 \frac{(\Theta e^N \cosh \epsilon)'}{\Theta e^N}  
 =-\frac{q}{\kappa \Theta \sqrt{f} } + \frac{\beta q \eta}{n\Theta}\left(\log \frac{\beta q}{nr^2}\right)'.
\ee
This result generalizes the Tolman temperature gradient relation to a steady heat flow system. 
In the small $q$ expansion, the first/second term on the right-hand side corresponds to the $O(q)/O(q^2)$ contribution.
Therefore, when $q =0$, this equation reproduce the Tolman relation~\eqref{Tolman}.

\item 
Finally, the $q^a$ part of the energy-momentum conservation~\eqref{eq2} becomes, for a steady state,
\bea
\hat q^b \mathcal{K}_b = \left(\sigma-\frac{\beta q^2}{n\Theta^2}\right) \frac{q}{\kappa} 
	+\alpha \sqrt{f}q \sinh\epsilon \left(\log \frac{\alpha q}{nr^2}\right)' .
\label{Kl2}
\eea
Here, the left-hand side of Eq.~\eqref{Kl2} is 
$$
\hat q^b\mathcal{K}_b=\sqrt{f} e^{-N} (\chi e^N \cosh \epsilon)' .
$$
Putting together, we get a generalization of the Klein's relation for a steady heat flow,
\be{Kl2-2}
 \frac{(\chi e^N \cosh \epsilon)'}{\chi e^N}  
 =\left(\sigma-\frac{\beta q^2}{n\Theta^2}\right)\frac{q}{\kappa \chi \sqrt{f} } + \frac{\alpha q \eta}{n\chi}\left(\log \frac{\alpha q}{nr^2}\right)'.
\ee
Putting $\chi = \Theta \mu$ to this equation and using Eq.~\eqref{Tol2-2}, we get 
\be{mu'-2}
\mu' =
\frac{q}{\sqrt{f}\kappa \Theta}\frac{\mu+\sigma -\frac{\beta q^2}{n\Theta^2} }
						{  \sqrt{1+\eta^2/n^2} } 
+ \frac{q \eta}{n\Theta \sqrt{1+\eta^2/n^2}}
	\left[\nu \left(\log \frac{q}{nr^2}\right)'+ \alpha'- \mu \beta'\right]
,
\ee
where we use $\alpha = \mu\beta +\nu$ and Eq.~\eqref{id2}.
The first term presents first order terms in $q$ and the second term contributes to higher orders.
When $q=0$, this equation reproduces the Klein's relation~\cite{Klein49}.
\end{enumerate}
Here, we need a few comments for the four equations~\eqref{id2}, \eqref{sigma':2}, \eqref{Tol2-2}, and \eqref{mu'-2}.
For the present steady state, we have only three independent variables $n$, $s$, and $q$. 
On the other hand, there are four equations.
This fact signifies that one of the four equations must be redundant.
At present, we cannot present a formal proof for it. 
However, we present reasonable evidence in the next section.

\section{A steady state for mild heat flow } \label{sec:IV}

In the presence of an intense diffusion of matter, it is hard to believe that a steady state will survive for long periods of time unless there is a giant source or sink of matter/heat at the center.
We describe the steady heat flow by using the four equations~\eqref{id2}, \eqref{sigma':2}, \eqref{Tol2-2}, and \eqref{mu'-2}, which are not simple in general. 
To get physical intuition for the effects of the heat, we consider a mild heat flow system satisfying 
\be{approx}
q \ll n \Theta,
\ee 
in other words, $q/\varepsilon \ll 1$.
With this approximation, the energy density and the pressure in Eq.~\eqref{rho P} satisfy $\bar \rho \approx \rho \approx -\Lambda$ and $\bar P \approx \Psi$ to the first order in $q$.
Then, from Eq.~\eqref{epsilon:q} and \eqref{id2}, we have $\epsilon = q/(\rho+\Psi) \ll 1$ and $\eta \ll n$.

\vspace{.3cm}
Let us observe the four equations to the first order in $q$. 
The particle creation equation~\eqref{id2} presents, 
\be{id1:1}
\eta = \frac{q}{ \Theta (\mu+\sigma)}+O(q^3) \approx \frac{J_\infty}{4\pi r^2 \sqrt{-g_{tt}}} ,
\ee
where $\approx$ denotes the same up to the first order in $q$ and the second equality comes from the invariant quantity, $J_\infty$ in Eq.~\eqref{id1}.
Note that the left/right-hand side of this equation has a thermodynamical/geometrical origin. 
It is interesting to compare the present result with the perturbational results around a thermal equilibrium of the form $\delta F \propto e^{\Gamma t + i k x}$, where $F$ denotes any thermodynamical quantity.
Here, positive $\Gamma$ signals the presence of instability. 
In the presence of a heat fluctuation $\delta q$, the particle creation equation relates the number density perturbation $\delta n$ to $\delta u_r$ (See Eq.~(67) in~\cite{LK2022}).
The relation comes from the time dependence of the perturbations.
On the other hand, when a non-vanishing heat flows, we find that a position-independent characteristic $J_\infty$ appears.
Note also that there are no $O(q^2)$ corrections.

The entropy creation equation~\eqref{sigma':2} presents a formula for the spatial gradient of the chemical potential to temperature ratio:
\be{sigma':3}
\mu'
=\frac{q (\mu+\sigma)}{\sqrt{f} \kappa \Theta} +O(q^2).
\ee
This equation~\eqref{sigma':3} determines how $\mu$ varies spatially because of the mild heat. 
The ratio increases with $r$ when the heat flows out, contrasting with the equilibrium result, $\mu'=0$.
When perturbing a thermal equilibrium mentioned above, the entropy creation equation presents a relation between the heat $\delta q$ and the variation of specific entropy $\delta\sigma$, which appears because of the time dependence of the perturbation.
Interestingly, the perturbative result for the energy-momentum conservation equation~\eqref{eq2} presents the relation between $\delta \mu$ and $\delta q$~(See Eq.~(72) in Ref.~\cite{LK2022}.), which is consistent with Eq.~\eqref{sigma':3}.

The relativistic analogy of the Cattaneo equation~\eqref{Tol2-2} becomes
\be{Tol:g}
\frac{\Theta'}{\Theta} + N' = -\frac{q}{\kappa \Theta\sqrt{f}}  +O(q^2)
	\approx -\frac{\mu+\sigma}{  4\pi r^2\sqrt{f} e^N} \frac{J_\infty}{\kappa},
\ee
where we use Eq.~\eqref{id1:1} in the last equality.
In thermal equilibrium with $q=0$, the equation reproduces the Tolman temperature gradient. 
In this sense, we interpret this equation as a generalization of the Tolman temperature gradient in the presence of mild heat.
For comparison, we define a redshifted-local temperature (RSLT) at $r$:
\be{T infty}
T_{\infty}(r) \equiv \sqrt{-g_{tt}(r)} \Theta(r) ,
\ee
where we assume $\lim_{r\to \infty} g_{tt}(r)=-1$ for an asymptotically flat spacetime.
Note that $T_\infty$ is neither the asymptotic temperature nor the local temperature of a system.
It is the temperature of a light radiated from a subsystem at $r$ measured in an asymptotic region. 
Only when $q=0$, $T_\infty$ is identical to the asymptotic temperature.
Integrating both sides of Eq.~\eqref{Tol:g}, we get
\be{Tol3}
T_{\infty} (r) = T_{\infty} (r_-)  -\frac1{\kappa}\int_{r_-}^r \frac{q(r') e^{N(r')}}{ \sqrt{f(r')}} dr' .
\ee
Note that when $q> 0$, $T(r) < T(r_-)$ for $r> r_-$ and vice versa. 
Therefore, we may say that heat flows from a higher RSLT region to a lower RSLT region, which result is consistent with our intuition that thermal equilibrium is characterized by RSLT.
Conversely, putting the formula for $q$ in Eq.~\eqref{id1:1} to the above equation and using the definition~\eqref{T infty} and $e^N= \sqrt{f} e^{-h/2}$, we get a differential equation for the RSLT,
\be{dT inf}
T_\infty'(r) = -\frac{1}{\kappa} \frac{J_\infty e^N \Theta(\mu+\sigma)}{4\pi r^2\sqrt{f} e^N} 
	= -\frac{J_\infty}{\kappa} \frac{(\mu+\sigma)e^{h/2} }{4\pi r^2f } T_\infty(r) .
\ee
This presents an interesting differential equation for $T_\infty$.
We discuss its consequences around an event horizon in the last section~\ref{sec:V}.
In the absence of a transverse perturbations, the perturbative result around a thermal equilibrium state relates $\delta \Theta$ with the heat $\delta q$ (See Eq.~(69) in~\cite{LK2022}), which is consistent with Eq.~\eqref{Tol:g}. 

As mentioned at the end of the previous section, the final equation~\eqref{mu'-2} originating from the energy-momentum conservation fails to present a new equation but gives a redundant equation with Eq.~\eqref{sigma':3} to the first order. 
Therefore, to this order, three independent equations~\eqref{id1:1}, \eqref{sigma':3}, and \eqref{Tol:g} determine the three independent variables $n$, $s$, and $q$.
It is natural to ask this redundancy of the equations holds even higher orders in $q$. 
We check this possibility up to seventh order in $q$ by using the Mathematica program.
Then, we find that the equations~\eqref{sigma':3} and \eqref{mu'-2} are identical up to the other two equations~\eqref{id2} and \eqref{Tol2-2}.  
Even though it is not a sufficient proof for the redundancy of the four equations~\eqref{id2}, \eqref{Tol2-2}, \eqref{sigma':2} and \eqref{mu'-2}, this fact and the presence of only three independent parameters provide convincing evidence.

\section{Summary and Discussions} \label{sec:V}

We have studied steady state of heat conduction in general relativity based on the two-fluid model of the variational approach for fluid dynamics. 
We adopted a Landau-Lifschitz (LL) coordinates instead of the Eckart frame, in which  observers move with fluid elements.
The reason for this choice is as follows: 
In comoving coordinates with matter, heat appears as a time-space element of the stress tensor, making the geometry inevitably time-dependent when the comoving vector $u^a$ generates the coordinate time.
This fact is against the existence of a timelike Killing vector for a steady state.
Therefore, we choose to use the LL frame for thermodynamic quantities, in which the energy flux vanishes to allow time-independent coordinate choice.
Since the heat conduction equations employ the Eckart frame, we have obtained a relationship between the two frames.
In doing so, we showed that the stability condition of thermal equilibrium found in the literature is simply the condition that the relative speed between heat and number fluxes is not faster than the light.
Then, we summarized the heat conduction equations, emphasizing the binormal equilibrium condition that imposes the Tolman temperature gradient along the perpendicular directions to the heat.
We also formulated the steady heat conduction equation for radial heat flow in a spherically symmetric spacetime.
We found a position-independent characteristic quantity,
\be{J inf}
 J_\infty \equiv \sqrt{-g_{tt}} J(r),
\ee
where $J(r) = -4\pi r^2 (n^a \hat j_a)$ with $(n^a \hat j_a)$, the diffusion of the number flux at $r$ with respect to the LL frame as in Eq.~\eqref{u v}.  
Here, the $4\pi r^2$ factor denotes the area of a constant $r$ surface.
Therefore, $-J(r)$ denotes that the total diffusion over spherical surface.
In this work, the momentum flow due to the diffusion cancels the momentum flow due to heat so that the total energy flux vanishes. 
The quantity $J_\infty$ characterizes the system with steady heat conduction in this sense.
We also wrote a generalized Tolman relation satisfied by the local temperature in the presence of steady heat conduction.

We have also studied systems with mild heat flux satisfying $q \ll n \Theta$ as an explicit example. 
In this case, the heat satisfies 
\be{q J}
\frac{q}{ \Theta(\mu+\sigma)}\approx \frac{J_\infty}{4\pi r^2 \sqrt{-g_{tt}}} ,
\ee
where $\Theta$, $\mu$, and $\sigma=s/n$ denote the local temperature, the chemical potential to temperature ratio, and the specific entropy, respectively.
We also found that the temperature satisfies a generalized Tolman relation, which allows us to define a red-shifted local temperature (RSLT) $T_\infty(r)$, corresponding to the asymptotic temperature for a thermal equilibrium system, as in Eq.~\eqref{Tol3}.
We have shown that heat flows from a higher RSLT region to a lower RSLT region.
In this sense, RSLT plays the same role as temperature for systems without gravity.

\vspace{.2cm}
With a black hole and Hawking radiation in mind, let us observe what happens to the heat around a horizon, where $f(r)$ goes to zero. 
Equation~\eqref{dT inf} presents an interesting information.  
Around the horizon, the thermodynamic parameter $\mu+\sigma$ and the metric function $e^{h/2}$ varies slowly with $r$.
Therefore, we may set them their horizon values and watch the behavior to get
\be{T' T}
\frac{T_\infty'}{T_\infty} \approx -\frac{J_\infty H}{ \kappa} 
	\frac{1/r_H}{r/r_H-1} ; \qquad 
H = \left(\frac{ (\mu+\sigma)e^{h/2} }{4\pi r } 
	\right)_{r= r_H},
\ee
where $r_H$ and $\kappa$  denote the horizon radius and the heat conductivity, respectively.
Integrating around the horizon, we get the behavior of RSLT around the horizon
\be{T hor}
T_\infty(r) \approx T_c\left( \frac{r}{r_H} -1 \right)^{-\frac{QH}{\kappa}} ,
\ee
where $T_c$ is an integration constant having a temperature dimension.
Now, we have two interesting results. 
First, when $J_\infty>0$ (heat is outgoing), the value of RSLT diverges at the horizon however small the $J_\infty$.
In other words, to avoid the divergence of RSLT, heat cannot get out of the horizon, which is a well known consequence of an event horizon.
Second, when $J_\infty< 0$ (heat is propagating inward), the value of RSLT goes to zero at the horizon however small the $|J_\infty|$.   
The local temperature of the system at the horizon is given by 
\be{Theta TH}
\Theta(r_H) = \lim_{r\to r_H}\frac{T_\infty(r)}{\sqrt{-g_{tt}}} \propto \left(\frac{r}{r_H}-1\right)^{\frac{-J_\infty H}{\kappa} -\frac12}.
\ee
When $-J_\infty H/\kappa \gtrless 1/2$, the local temperature goes to zero/diverges at the horizon.  
If we want the local temperature of a system at the horizon is finite as one usually expects, we have
\be{J inf hor}
J_\infty = -\frac{\kappa}{2H}.
\ee
In other words, the heat should be absorbed into the horizon with the specific value of $J_\infty$.
A caveat should be given here. 
Around the horizon, the value of $q$ increases making the linear level analysis in Sec.~\ref{sec:IV} invalid. 
However, we can still consider small enough $J_\infty$ to approach the horizon indefinitely.

\vspace{.2cm}
In this work, we have assumed that the steady thermal system is located in a compact region, $r_- \leq r \leq r_+$, to avoid dynamic behaviors of the system. 
It is natural to ask what happens when $r_- $ approaches zero or $r_+ \to \infty$. 
As shown in the heat formula~\eqref{id1:1}, the intensity of heat diverges inverse quadratically with $r$ as $r\to 0$.
Its integration over the spherical surface of constant $r$ provides a finite energy transfer, $ \Theta(\mu+\sigma)J_\infty/\sqrt{-g_{tt}}$. 
Therefore, this limit presumes the presence of a very hot and small condensed heat reservoir.
In the absence of such a condensed heat reservoir, the system cannot be in a steady state. 
Then, one should take into account dynamical behaviors such as particle creation at the center.
Note also that for intense heat flux, one cannot ignore the non-linear terms in the equation of motions.
In that case, one needs to know how the master function or the energy density depends on heat to identify the $\beta$ term. 
Additional analysis will be required to notify the non-linear behaviors. 
The other regime to check is the asymptotic region, $r_+ \to \infty$.
In the region, the heat decreases as $1/r^2$ because $\sqrt{-g_{tt}} \to 1$ and $\Theta \to T_\infty$.
The integration over the constant $r$ surface presents a mildly varying finite energy transfer $\sim(\mu+\sigma) T_\infty J_\infty$.

\vspace{.2cm}
We also presented convincing arguments that one of the heat flow equations~\eqref{id2}, \eqref{Tol2-2}, \eqref{sigma':2} and \eqref{mu'-2} becomes redundant because of the steady heat flow condition. 
Formal proof for the redundancy is anticipated.
Because the stability of a thermal equilibrium state is known, the steady system with mild heat flow is sure to be stable with appropriate conditions. 
However, the stability of systems with intense heat is still questionable and requires further studies. 

%
\section*{Acknowledgment}
This work was supported by the National Research Foundation of Korea grants funded by the Korea government RS-2023-00208047.
The author thanks to Dr. Youngone Lee for helpful discussions.

\appendix
\section{Calculations } \label{app:calc}
In this appendix, we present a few details of the calculations. 
We first write the derivative of $\log (q/n\Theta)$ in terms of $\tanh 2\epsilon$ by using Eqs.~\eqref{epsilon q} and \eqref{n ep}:
\bea
 \left(\log \frac{q}{n\Theta}\right)' &=&  \left(\log \frac{\varepsilon}{n\Theta}\right)' 
 	+ \left(\log( \tanh 2\epsilon)\right)' 
	=  \left(\log \frac{\varepsilon}{n\Theta}\right)' 
 	+ \left(\log \frac{2\tanh \epsilon}{1+ \tanh^2\epsilon} \right)' \nn \\
 &=&  \left(\log \frac{\varepsilon}{n\Theta}\right)' 
 	 +\left(\log (n \sinh \epsilon)\right)'
	 + \frac12\left(\log [n^2+(n\sinh\epsilon)^2] \right)' 
	 - \left(\log [n^2+ 2(n\sinh\epsilon)^2]\right)' \nn \\
&=&  \left(\log \frac{\varepsilon}{n\Theta}\right)' 
 	 -(N'+\frac2r) 
	 +\left[1+ \frac1{\frac{\eta^2}{n^2}}- \frac{2}{1+\frac{2\eta^2}{n^2}} \right]\frac{n'}{n} 	
	 +\left[\frac1{1+\frac{\eta^2}{n^2}}-\frac{4}{1+\frac{2\eta^2}{n^2}}\right]\frac{\eta^2}{n^2}  \left(\log (n\sinh\epsilon)\right)' \nn \\
&=& \left(\log \frac{\varepsilon}{n\Theta}\right)' 
	 +\left(-1+\frac{3 + 2\frac{\eta^2}{n^2}}{(1+\frac{\eta^2}{n^2})(1+\frac{2\eta^2}{n^2})} 
		 \frac{\eta^2}{n^2}\right) \left( \frac{n'}{n} 	
	 +N'+\frac2r\right) . \label{app:qT}
\eea
We next write the derivative of $\epsilon$ in terms of $\eta$ and other derivatives:
\be{app:chep'}
(\log \cosh\epsilon)' = \frac12 \left(\log [n^2+(n\sinh\epsilon)^2]\right)'- \frac{n'}{n}
= - \left(N'+\frac2r +\frac{n'}{n}\right) \frac{\eta^2/n^2}{1+\eta^2/n^2} . 
\ee

Now, we write the Christoffel symbol for the metric~\eqref{metric1}.
The non-vanishing components of the Christoffel symbol are
\bea
&& \Gamma^0_{01} = N', 
\qquad
\Gamma^1_{11} = -\frac{f'}{2f}, \qquad
\Gamma^1_{00} = N' f e^{2N} , \nn \\
&&\Gamma^1_{22} = -r f, \qquad
\Gamma^{1}_{33} = - r f  \sin^2\theta, \qquad \Gamma^2_{12} = \frac1r, \qquad
\Gamma^2_{33} = -\sin \theta \cos\theta , \quad
\Gamma^3_{13} = \frac1r, \qquad
\Gamma^3_{23} = \cot \theta .
\eea
Then, let us calculate the proper-time derivative of the heat, $dq_a/d\tau$:
\bea
\frac{d q_a}{d\tau} &=& u^b \nabla_b q_a
 =  (\cosh \epsilon \, v^b- \sinh\epsilon \,\hat j^b) \nabla_b  (-\sinh \epsilon \, q v_a + q \cosh \epsilon \, \hat j_a) \nn \\
&\simeq&- \sinh\epsilon \, \hat j^b \partial_b 
	(-\sinh \epsilon \, q v_a + q \cosh \epsilon \, \hat j_a)
 + \sinh\epsilon \, \hat j^b \Gamma^d_{ba}  
 	 (-\sinh \epsilon \, q v_d + q \cosh \epsilon \, \hat j_d)\nn \\
 &\approx & \sqrt{f}q \sinh \epsilon 
 	\left[-\frac{q'}{q} \, \hat q_a 
	+\epsilon'\, u_a \right] .
	  \label{app:dq/dtau}
\eea
Here, we use the steady state condition~\eqref{stationary} to remove terms of the form $v^a \nabla_a \epsilon$.

\vspace{.2cm}
We next calculate the covariant derivative of $u_a$ for the spherically symmetric metric~\eqref{metric1},
\bea
\nabla_a u_b &=& \partial_a u_b - \Gamma_{ab}^d u_d 
= \partial_a(\cosh \epsilon \, v_b - \sinh \epsilon \, \hat j_b) 
	- (\cosh \epsilon \, v_d -\sinh \epsilon \hat j_d) \Gamma^d_{ab} \nn \\
&=& \delta_a^1 \epsilon' (\sinh \epsilon \, v_b -\cosh \epsilon \, \hat j_b) + \cosh \epsilon \, \delta_a^1 \delta_b^0 (-e^N)'
	- \sinh \epsilon \, \delta_a^1 \delta_b^1 (f^{-1/2})'
	+ \cosh \epsilon \, e^N (\delta_a^0\delta_b^1+\delta_a^1\delta_b^0)\Gamma^0_{10} 
	+  \frac{\sinh\epsilon }{\sqrt{f}} \Gamma^1_{ab} \nn \\
&=& (-\delta_a^1 \epsilon'  
	+\delta_a^0 \sqrt{f} e^N N' )\hat q_b
- \sinh\epsilon \, r\sqrt{f} 
		\left[\delta_a^2\delta_b^2   
		+\delta_a^3\delta_b^3  \sin^2\theta \right] .
\eea
Contracting the covariant derivative with $q^b$, we get
\bea \label{app:d u q}
(\nabla_a u_b)q^b &=& (-\delta_a^1 \epsilon'  +\delta_a^0 \sqrt{f} e^N N' )q
 = -[(N' \cosh\epsilon +\epsilon' \sinh \epsilon) qu_a 
 	+ (N' \sinh \epsilon + \epsilon' \cosh\epsilon) q_a] \sqrt{f} 
 , \nn \\
q^c \nabla_c u_b 
	&=& q^c(-\delta_c^1 \epsilon'  +\delta_c^0 \sqrt{f} e^N N' )\hat q_b
	=  -(\epsilon' \cosh \epsilon  +  N' \sinh \epsilon )\sqrt{f} q_b .
\eea
Therefore, 
\be{app:q d u}
q^c(\nabla_c u_b -\nabla_b u_c) 
=  (N' \cosh \epsilon + \epsilon' \sinh \epsilon) \sqrt{f} q u_a .
\ee

\vspace{.2cm}
Next, we calculate the Tolman vector $\mathcal{T}_a \equiv \frac{d\Theta u_a}{d\tau}+ \nabla_a \Theta$ in a static spherically symmetric coordinates. 
For this purpose, we first calculate $q^a du_a/d\tau$ and $d\Theta/d\tau$.
\bea
\frac{d u_a}{d\tau} = u^c\nabla_c u_a &=& u^c (\partial_c u_a -\Gamma^b_{ca} u_b)
= (\cosh \epsilon \, v^c -\sinh \epsilon \, \hat j^c) \partial_c (\cosh \epsilon \, v_a -\sinh \epsilon \, \hat j_a) \nn \\
&& - 	(\cosh \epsilon \, v^c -\sinh \epsilon \, \hat j^c) (\cosh \epsilon \, v_b -\sinh \epsilon \, \hat j_b) \Gamma^b_{ca} \nn \\
&\simeq&
 \sqrt{f} (\hat q_a) e^{-N}( e^N\cosh\epsilon)' .
\eea
In addition, we get
$$
\frac{d\Theta}{d\tau} = u^a \nabla_a \Theta 
= (\cosh \epsilon \, v^a - \sinh \epsilon \, \hat j^a) \nabla_a \Theta
\simeq -(\sinh \epsilon) \sqrt{f} \Theta'.
$$
Therefore, the Tolman vector becomes
\be{Tolman vector}
\mathcal{T}_a =  -(\sinh \epsilon) \sqrt{f} \Theta'u_a 
	+ \sqrt{f} (\hat q_a) [\epsilon'\sinh \epsilon +N' \cosh \epsilon] \Theta
	+\nabla_a \Theta .
\ee
Contracting the Tolman vector with $\hat q^a$, we get 
\be{Tolman q}
\hat q^a \mathcal{T}_a = 
	 \sqrt{f}  [\epsilon'\sinh \epsilon +N' \cosh \epsilon]
	 + (\cosh \epsilon \,\hat j^a - \sinh\epsilon \, v^a)\nabla_a\Theta
\simeq \sqrt{f} e^{-N} ( e^N \cosh\epsilon \,\Theta)' .
\ee


\end{document}